\documentclass[epjST]{svjour}
\usepackage{graphics}
\usepackage{graphicx}
\usepackage{amsmath}
\usepackage{amssymb}
\usepackage{textcomp}
\begin{document}
\title{Quantum many-body effects on Rydberg excitons in cuprous oxide}
\author{D. Semkat\inst{1}\fnmsep\thanks{\email{dirk.semkat@uni-greifswald.de}} \and H. Fehske\inst{1} \and H. Stolz\inst{2}}
\institute{
Institut f\"ur Physik, Ernst-Moritz-Arndt-Universit\"at Greifswald, Felix-Hausdorff-Str.\ 6, 17489 Greifswald, Germany
\and
Institut f\"ur Physik, Universit\"at Rostock, Albert-Einstein-Str.\ 23-24, 18059 Rostock, Germany
}
\abstract{
We investigate quantum many-body effects on Rydberg excitons in cuprous oxide induced by the surrounding electron-hole plasma. Line shifts and widths are calculated by full diagonalisation of the plasma Hamiltonian and compared to results in first order perturbation theory, and the oscillator strength of the exciton lines is analysed.
} 
\maketitle
\section{Introduction}
\label{intro}

Since their first experimental confirmation in cuprous oxide (Cu$_2$O) \cite{nature2014}, Rydberg excitons, i.e., excitons in states with high principal quantum numbers, have proven to be fascinating quantum objects with interesting properties (signatures of quantum coherence \cite{gruenwald2016}, occurrence of quantum chaos \cite{assmann2016}, to name only two examples).
Despite the great similarity of these excitonic states to Rydberg atoms, there have been several features in the absorption spectrum which cannot be explained by a simple atomic description and which are caused by pecularities of the solid state environment.
Recently it has been shown that a surrounding plasma of free carriers (electrons and holes) has a significant influence on the absorption spectrum, in particular on the position of the band edge, even though its density is very low \cite{heckoetter2018,semkat2019}. Moreover, residual charged impurities are expected to influence the spectrum in a very similar manner, causing in addition an exponential tail of the band absorption into the gap \cite{krueger-unpub}. Both scenarios co-occur and superimpose effects arising from the direct exciton-exciton interaction (Rydberg blockade).

In the present paper, we build up on a quantum many-body approach to the behaviour of bound (exciton) states and the band edge in a surrounding electron-hole plasma \cite{seidel1995,arndt1996,semkat2019}. The focus lies on the one hand on the comparison of the excitonic line shifts obtained by full diagonalisation of the plasma Hamiltonian to results in first order perturbation theory \cite{semkat2019}.
On the other hand, we look at the change of the oscillator strength and the damping of the Rydberg lines, aiming at answering the question whether and to what extent the observed bleaching of the lines before they merge with the band absorption and their linewidths are caused by quantum many-body effects.

\section{Theoretical approach}

\subsection{Exciton states}

We start from the Hamiltonian of an electron-hole pair in a surrounding of charge carriers, $\mathcal{H}=\mathcal{H}_0+\mathcal{H}'$ with
\begin{eqnarray}\label{hamiltonian0}
&&\mathcal{H}_0\psi(\mathbf{k},\omega)=\left(\frac{\hbar^2k^2}{2m_{\rm r}}-\hbar\omega\right)\psi(\mathbf{k},\omega)
-\int\frac{\mathrm{d}^3k'}{(2\pi)^3}V_{\rm eh}(|\mathbf{k}-\mathbf{k}'|)\psi(\mathbf{k}',\omega)
\end{eqnarray}
describing the isolated two-particle problem and the many-body (plasma) induced part $\mathcal{H}'$ which reads \cite{5maenner,gruenesbuch}
\begin{eqnarray}\label{hamiltonian}
&&\mathcal{H}'\psi(\mathbf{k},\omega)=-\int\frac{\mathrm{d}^3k'}{(2\pi)^3}
\left\{
V_{\rm eh}(|\mathbf{k}-\mathbf{k}'|)\left[f_{\rm e}(\mathbf{k})+f_{\rm h}(-\mathbf{k})\right]
\psi(\mathbf{k}',\omega)\right.\\
&&\vspace*{10ex}
-V_{\rm eh}(|\mathbf{k}-\mathbf{k}'|)\left[f_{\rm e}(\mathbf{k}')+f_{\rm h}(-\mathbf{k}')\right]\psi(\mathbf{k},\omega)
\left.+\Delta V_{\rm eh}^{\rm eff}(\mathbf{k},\mathbf{k}',\omega)\left[\psi(\mathbf{k}',\omega)-\psi(\mathbf{k},\omega)\right]
\right\}\nonumber
\end{eqnarray}
with $\psi(\mathbf{k},\omega)$ being the excitonic wave function and $\Delta V_{\rm eh}^{\rm eff}$ the effective potential,
\begin{eqnarray}\label{veff}
&&\Delta V_{\rm eh}^{\rm eff}(\mathbf{k},\mathbf{k}',\omega)=V_{\rm eh}(|\mathbf{k}-\mathbf{k}'|)
\int\limits_{-\infty}^{\infty}\frac{\mathrm{d}\bar{\omega}}{\pi}\,
\mathrm{Im}\,\varepsilon^{-1}(\mathbf{k}-\mathbf{k}',\bar{\omega}+\mathrm{i}\epsilon)\\
&&\times\left\{
\frac{n_{\rm B}(\bar{\omega})+1}{\hbar\omega+\mathrm{i}\epsilon-\hbar\bar{\omega}-E_{\rm e}(\mathbf{k}')-E_{\rm h}(-\mathbf{k})}\right.
+\left.
\frac{n_{\rm B}(\bar{\omega})+1}{\hbar\omega+\mathrm{i}\epsilon-\hbar\bar{\omega}-E_{\rm e}(\mathbf{k})-E_{\rm h}(-\mathbf{k}')}
\right\}\nonumber
\end{eqnarray}
with Coulomb potential $V_{ab}(k)=e_ae_b/(\varepsilon_0\varepsilon_{\rm b}k^2)$ ($a,b=\mathrm{e,h}$ and $\varepsilon_{\rm b}$ is the background dielectric constant), dielectric function $\varepsilon(\mathbf{k},\omega)$, Bose distribution $n_{\rm B}(\omega)=\left(\mathrm{exp}\left[\hbar\omega/(k_{\rm B}T)\right]-1\right)^{-1}$, and reduced mass $m_{\rm r}=m_{\rm e}m_{\rm h}/(m_{\rm e}+m_{\rm h})$. We assumed zero center-of-mass momentum of the electron-hole pair and weak degeneracy, i.e., not too high densities $n_a$ and not too low temperatures $T_a$ so that $n_a\Lambda_a^3\ll 1$, where $\Lambda_a$ is the thermal deBroglie wave length, $\Lambda_a^2=2\pi\hbar^2/(m_ak_{\rm B}T_a)$.

The exciton wave functions are given by the usual factorisation into radial part and spherical harmonics, $\left<nlm|\mathbf{k}\right>=\psi_{nlm}(\mathbf{k})=\phi_{nl}(k)Y_{lm}(\vartheta,\varphi)$, where the radial functions $\phi_{nl}$ can be expanded into the radial eigenfunctions of the unperturbed electron-hole bound states, i.e., the Coulomb eigenfunctions,
\begin{equation}\label{phi}
\phi_{nl}(k)=\sum\limits_{n'}c_{n,n'}^{l}\phi_{n'l}^{\rm C}(k)\,.
\end{equation}
The electron-hole Hamiltonian has to be represented in that basis. Then the plasma-perturbed exciton energies are given by solving the eigenvalue problem
\begin{equation}\label{eigen}
E_{nlm}\,|nlm\rangle=\mathcal{H}|nlm\rangle\,,
\end{equation}
i.e., by diagonalisation of the Hamiltonian
\begin{eqnarray}\label{DeltaE}
\left<nlm|\mathcal{H}|n'l'm'\right>&=&E_{nl}^{(0)}\,\delta_{nn'}\delta_{ll'}\delta_{mm'}+\left<nlm|\mathcal{H}'|n'l'm'\right>\\
\mbox{with}\quad\left<nlm|\mathcal{H}'|n'l'm'\right>&=&
\int\mathrm{d}^3k\,\psi_{nlm}^*(\mathbf{k},\omega)\mathcal{H}'\psi_{n'l'm'}(\mathbf{k},\omega)\equiv\Delta \mathcal{H}_{nn'}^{l}\,\delta_{ll'}\delta_{mm'}
\end{eqnarray}
yielding $E_{nlm}=E_{nl}=E_{nl}^{(0)}+\Delta E_{nl}$.
Here, $\Delta \mathcal{H}_{nn'}^{l}$, corresponding to the structure of $\mathcal{H}'$, consists of four physically distinct contributions \cite{5maenner,gruenesbuch}, two static ones (Pauli blocking and Hartree--Fock self-energy; first and second summands on the r.h.s. of Eq.\ (\ref{hamiltonian})), and two dynamic ones (dynamical self-energy correction and dynamically screened effective potential, both contained in the third summand). Thus, $\Delta \mathcal{H}_{nn'}^{l}=\Delta \mathcal{H}_{nn'}^{l;\mathrm{stat}}+\Delta \mathcal{H}_{nn'}^{l;\mathrm{dyn}}$ with \cite{semkat2019}
\begin{eqnarray}\label{deltaepbf}
&&\Delta \mathcal{H}_{nn'}^{l;\mathrm{stat}}=\frac{e^2}{\varepsilon_0\varepsilon_{\rm b}}\frac{1}{(2\pi)^2}\int\limits_0^{\infty}\mathrm{d}k\int\limits_0^{\infty}\mathrm{d}k'\,kk'\left[f_{\rm e}(k)+f_{\rm h}(k)\right]\phi_{n'l}^{\rm C}(k')\nonumber\\
&&\times\left\{ \phi_{nl}^{{\rm C}*}(k)Q_l\left(\frac{k^2+k'^2}{2kk'}\right)-\phi_{nl}^{{\rm C}*}(k')Q_0\left(\frac{k^2+k'^2}{2kk'}\right)\right\}\,,
\end{eqnarray}
where $P_l$ is the Legendre polynomial and $Q_l$ is the Legendre function of the second kind,
$Q_l(z)=\frac{1}{2}\int_{-1}^1\mathrm{d}t\,P_l(t)/(z-t)$,
and
\begin{eqnarray}\label{deltaedsp}
&&\Delta \mathcal{H}_{nn'}^{l;\mathrm{dyn}}=
-\frac{e^2}{\varepsilon_0\varepsilon_{\rm b}}\frac{1}{(2\pi)^2}\int\limits_0^{\infty}\mathrm{d}k\int\limits_0^{\infty}\mathrm{d}k'\,kk'\,
\int\limits_{-1}^1\mathrm{d}t\,\frac{1}{\frac{k^2+k'^2}{2kk'}-t}\nonumber\\
&&\times\left\{ \phi_{nl}^{{\rm C}*}(k)\phi_{n'l}^{\rm C}(k')P_l(t)-\frac{1}{2}\left[\phi_{nl}^{{\rm C}*}(k)\phi_{n'l}^{\rm C}(k)+\phi_{nl}^{{\rm C}*}(k')\phi_{n'l}^{\rm C}(k')\right]\right\}\nonumber\\
&&\times\Bigg\{
[1+n_{\rm B}(\omega_0)]
\left[\mathrm{Re}\,\varepsilon^{-1}(\sqrt{k^2+k'^2-2kk't},\omega_0)-1\right]\nonumber\\
&&-\frac{k_{\rm B}T}{\hbar\omega_0}
\left[\varepsilon^{-1}(\sqrt{k^2+k'^2-2kk't},0)-1\right]
-\frac{4\hbar\omega_0}{k_{\rm B}T}\sum\limits_{j=1}^{\infty}\frac{\varepsilon^{-1}(k,\mathrm{i}2\pi jk_{\rm B}T/\hbar)-1}
{\left(\frac{\hbar\omega_0}{k_{\rm B}T}\right)^2+(2\pi j)^2}\Bigg\}
\end{eqnarray}
\begin{eqnarray}
\mbox{with}\qquad\hbar\omega_0=\hbar\omega_0(k,k')=E_{nl}-\frac{\hbar^2k^2}{2m_{\rm e}}-\frac{\hbar^2k'^2}{2m_{\rm h}}\,.
\end{eqnarray}
As in Ref.\ \cite{semkat2019}, the dielectric function $\varepsilon(\mathbf{k},\omega)$ will be used in the nondegenerate limit of the ``random phase approximation'' (RPA) \cite{fehr1994}.

Note that the present approach does not account for higher (e.g. three- and four-particle) correlations so that higher bound states like trions (charged excitons) and biexcitons are excluded. However, they are not expected to play a role in Cu$_2$O anyway.

\subsection{Oscillator strengths}

While the vanishing of the highest Rydberg exciton lines with increasing pump power has been identified as a many-body effect (Mott effect) already in Ref.\ \cite{heckoetter2018}, another puzzle formulated in the pioneering work \cite{nature2014} remained unsolved even in the many-body theoretical analysis \cite{semkat2019}. It concerns the gradual bleaching of the exciton lines before they finally vanish in the band. The oscillator strength of the lines is proportional to the square of the derivative of the excitonic wave function at $r=0$. For an unperturbed exciton one obtains \cite{elliott1957}
\begin{equation}
f_{nl}^{(0)}\propto a_{\rm X}^{-5}\frac{n^2-1}{n^5}\,.
\end{equation}
However, the analysis of the spectra revealed a deviation from the expected $n^{-3}$-dependence for higher $n$ \cite{nature2014}, i.e., an unidentified bleaching mechanism.

In order to clarify whether that bleaching is caused by quantum many-body effects, we differentiate the plasma-perturbed wave function, Eq.\ (\ref{phi}). One obtains for the relative change of the oscillator strength
\begin{equation}
\frac{f_{nl}}{f_{nl}^{(0)}}=\frac{\sum\limits_{n'}c_{n,n'}^{l}\frac{n'^2-1}{n'^5}}{\frac{n^2-1}{n^5}}\,.
\end{equation}

\subsection{Line widths}

A further property of the Rydberg exciton lines, which has not been considered from the many-body theoretical point of view so far, is the line width. This quantity is a measure for the life time of the exciton states.
While the energy of an exciton state is given by the eigenvalue of the real (Hermitian) part of the Hamiltonian [cf.\ Eq.\ (\ref{eigen})], its width is determined by the corresponding imaginary part \cite{seidel1995}
\begin{eqnarray}\label{imaghstrich}
&&\mathrm{Im}\,\mathcal{H}'(\mathbf{k},\mathbf{k'},\omega)=-\int\mathrm{d}^3q\,
V_{\rm eh}(q)
\left[\delta(\mathbf{k'}-\mathbf{k})-\delta(\mathbf{q}-(\mathbf{k'}-\mathbf{k}))\right]
\\
&&\times
\left\{
\left(n_{\rm B}(\omega-(E_{\rm h}(\mathbf{k})+E_{\rm e}(\mathbf{k}'))/\hbar)+1\right)
\,\mathrm{Im}\,\varepsilon^{-1}(\mathbf{q},\bar{\omega}-(E_{\rm h}(\mathbf{k})+E_{\rm e}(\mathbf{k}'))/\hbar+\mathrm{i}\epsilon)
\right.
\nonumber\\
&&\vspace*{10ex}
+\left.
\left(n_{\rm B}(\omega-(E_{\rm e}(\mathbf{k})+E_{\rm h}(\mathbf{k}'))/\hbar)+1\right)
\,\mathrm{Im}\,\varepsilon^{-1}(\mathbf{q},\bar{\omega}-(E_{\rm e}(\mathbf{k})+E_{\rm h}(\mathbf{k}'))/\hbar+\mathrm{i}\epsilon)
\right\}\,,\nonumber
\end{eqnarray}
more precisely, by its expectation value with the eigenstates determined by Eq.\ (\ref{eigen}) for $\hbar\omega$ being the corresponding eigenvalue.

\section{Results}

\begin{figure}[hb]%
\begin{center}
\includegraphics*[width=0.32\textwidth]{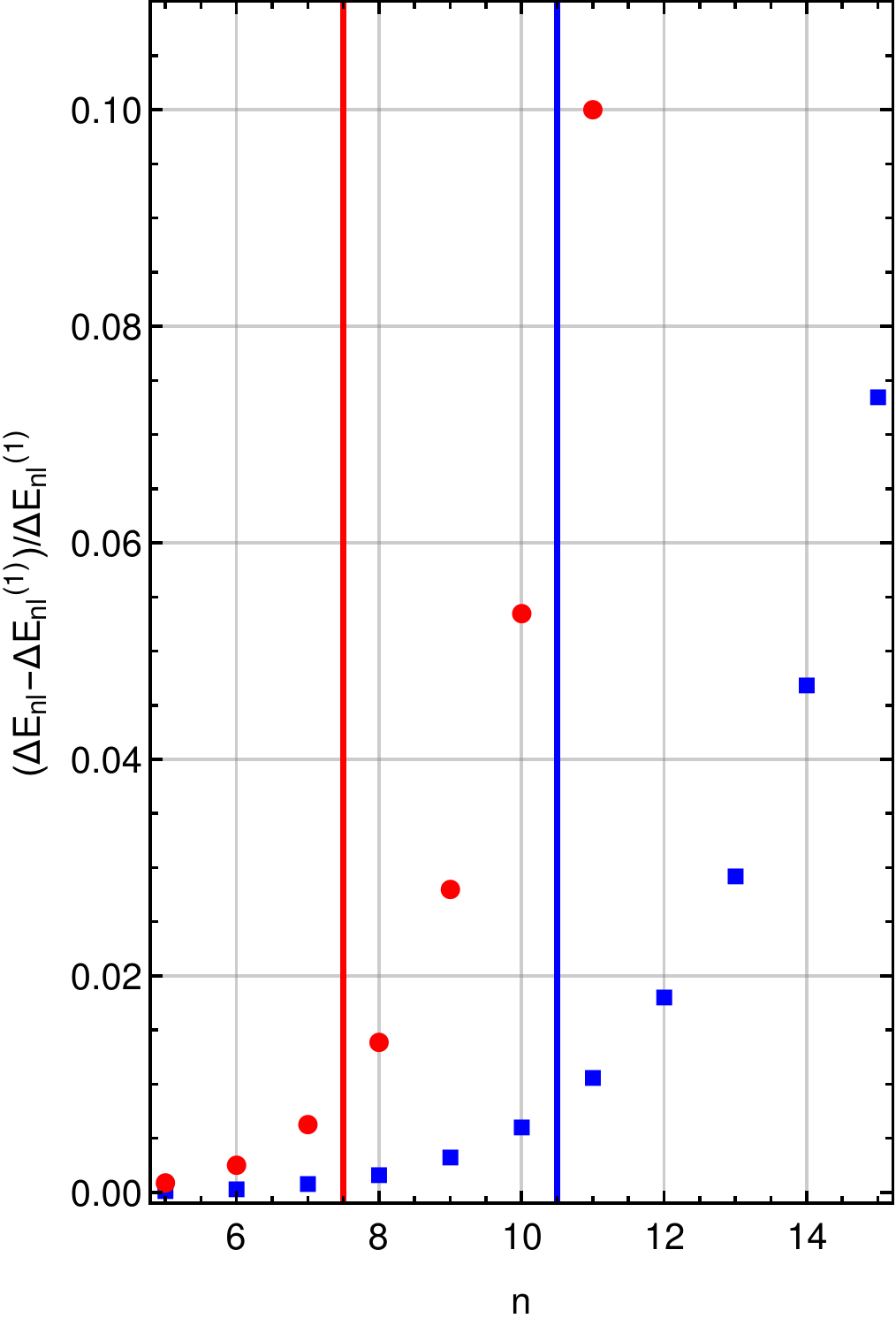}\hspace*{0.1cm}
\includegraphics*[width=0.32\textwidth]{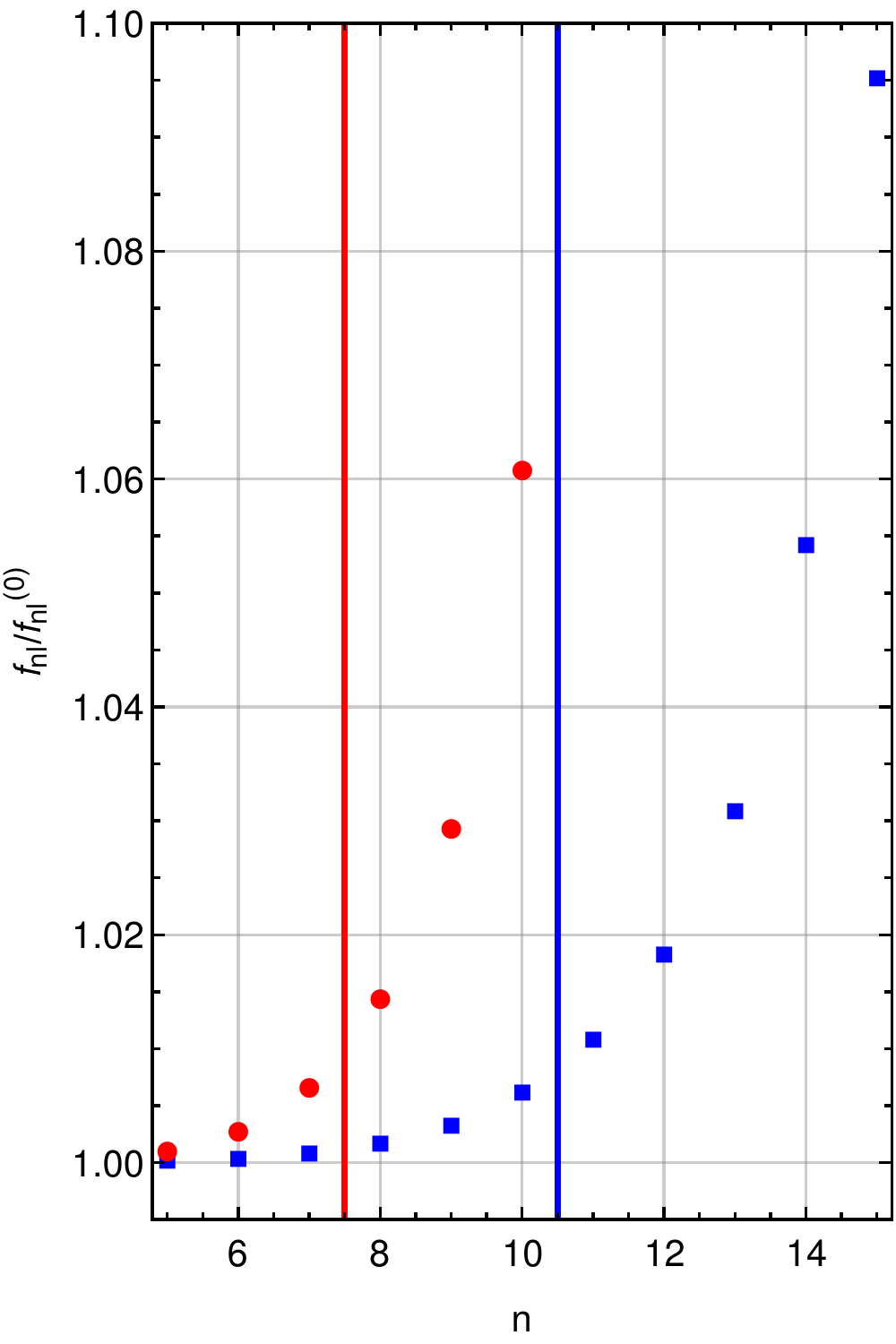}\hspace*{0.1cm}
\includegraphics*[width=0.32\textwidth]{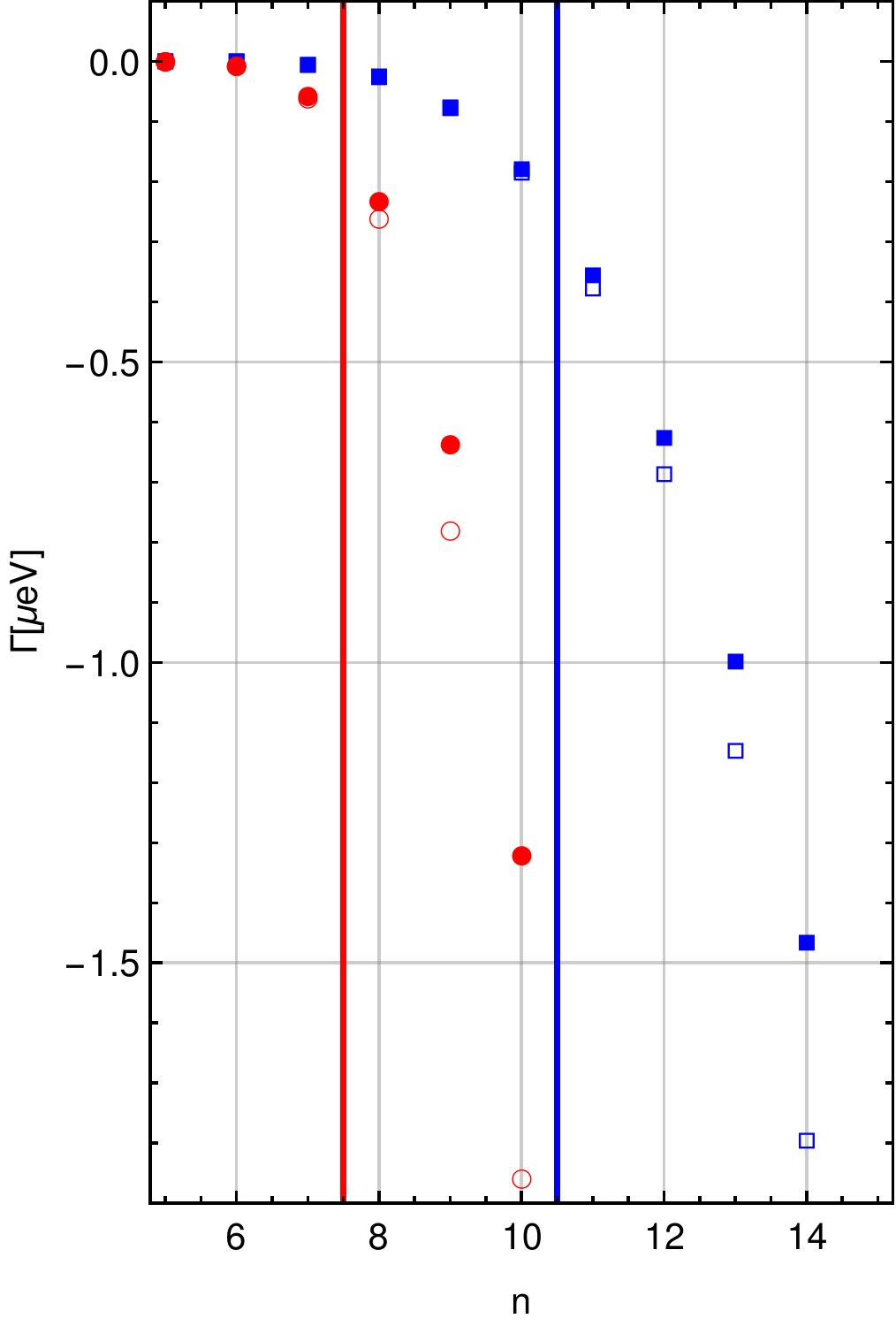}
\caption{(a) Relative difference between exciton level shifts obtained by full diagonalisation of the electron-hole pair Hamiltonian and those in first order perturbation theory;
(b) oscillator strengths of the exciton states in relation to the unperturbed case;
(c) line widths of the exciton states obtained by full diagonalisation of the electron-hole pair Hamiltonian (filled symbols) and in first order perturbation theory (empty symbols); each \textit{vs.}\ principal quantum number $n$ for a temperature of $T=10$ K and two carrier densities: $\rho_{\rm eh}=10^{11}$ cm$^{-3}$ (blue squares) and $\rho_{\rm eh}=10^{12}$ cm$^{-3}$ (red circles).
Lines right from the respective vertical bars vanish in the band due to the Mott effect.
}
\label{fig:shifts-fzuf0-gamma}
\end{center}
\end{figure}

Numerical results for the quantities introduced in the previous section are depicted in Fig.\ \ref{fig:shifts-fzuf0-gamma}.
The left panel shows the relative difference between the energy shift $\Delta E_{nl}$ of the exciton levels according to Eqs.\ (\ref{eigen})--(\ref{deltaedsp}) and the corresponding result in first order perturbation theory $\Delta E_{nl}^{(1)}$, see Ref.\ \cite{semkat2019}, for two different electron-hole densities.
Up to quite large $n$, the first order perturbation results fit the shifts obtained by full diagonalisation of the Hamiltonian well. Deviations occur only for higher $n$. They are more pronounced for the higher density. However, for a given temperature and density, the Mott effect \cite{gruenesbuch} causes a maximum achievable principal quantum number $n_{\rm max}$ \cite{heckoetter2018,semkat2019}, above which the exciton states merge with the continuum states in the conduction band. This is denoted by the vertical bars in all panels of Fig.\ \ref{fig:shifts-fzuf0-gamma}. Obviously, for all $n\le n_{\rm max}$, first order perturbation theory gives a very good approximation for the energy shifts. This is in agreement with corresponding experiences in the many-body physics of hydrogen plasmas \cite{fehr1994}.

The relative change of the oscillator strength is shown in the central panel for two chosen carrier densities \textit{vs.} principal quantum number. Obviously, always $f_{nl}/f_{nl}^{(0)}>1$, i.e., instead of a reduction of the lines, we find a slight enhancement by $\lesssim 1 \%$, thus an (although very weak) brightening of the lines. The plasma does not cause a line bleaching but even counteracts it! Therefore, many-body effects can be ruled out as the observed bleaching mechanism.
An enhancement, however, has not been found for all angular momentum states. For the ground ($1S$) state, one obtains $f_{nl}/f_{nl}^{(0)}\approx 0.98$ for $T=10$ K and $\rho_{\rm eh}=10^{18}$ cm$^{-3}$ (just below the Mott density), i.e., a weak bleaching of the line in qualitative agreement with earlier works \cite{zimmermann1988}.

The right panel shows $\Gamma$ for two densities \textit{vs.} principal quantum number $n$. Obviously, the plasma-induced line width is very small; it exceeds 1 \textmu eV only for lines well beyond the Mott boundary. Only in that region deviations between full calculation and first order perturbation theory become visible. The latter one even overestimates $\Gamma$. Plasma-induced many-body effects, thus, do not contribute significantly to the measured line widths \cite{nature2014,heckoetter2018}.

\section{Conclusions and outlook}

Based on a recently published quantum many-body approach to the behaviour of Rydberg exciton states in a surrounding electron-hole plasma \cite{semkat2019}, we improved the theoretical analysis to a full solution of the plasma-perturbed excitonic eigenvalue problem.  A comparison of the obtained excitonic line shifts to previous results in first order perturbation theory \cite{semkat2019} revealed a quite far-reaching agreement, showing that the latter provide, for the low plasma densities found in the experiments, a very good approximation to the full diagonalisation. This confirms similar results obtained for hydrogen plasmas \cite{seidel1995}.

For the line width of the exciton states, a similar agreement between full calculation and first order perturbation theory has been found. However, while the magnitude of the line shifts reproduces the experimental findings very well \cite{semkat2019}, the calculated widths are only of the order of a few hundred nanoelectronvolts and, thus, much smaller than those found in the experimental spectra. The actual line widths, therefore, must be caused by other physical mechanisms, e.g., interaction with phonons \cite{stolz2018}.

The oscillator strength of the exciton lines, calculated from the plasma-perturbed eigenstates, is slightly enhanced compared to the unperturbed case by $\lesssim 1 \%$. Thus, the observed bleaching of the lines before they merge with the band absorption cannot be caused by quantum many-body effects.  A very probable reason for the bleaching seems to be instead the action of spatially inhomogeneous fields induced by residual charged impurities \cite{krueger-unpub}.

In order to get a comprehensive picture of the action of the solid-state environment on Rydberg exciton states and the band edge, an in-depth analysis of measured transmission spectra is necessary, combined with a detailed comparison of experimental results with theoretical predictions. This is the subject of ongoing work.

%

\begin{acknowledgement}

We wish to thank S. O. Kr\"uger and W.-D. Kraeft (Rostock) for many fruitful discussions.
D.\ S.\ gratefully acknowledges support by the Deutsche Forschungsgemeinschaft (project number SE 2885/1-1).\\
D.\ S.\ developed the general concept and performed the calculations. All authors contributed equally to the discussion of the results and to the text of the manuscript.

\end{acknowledgement}

\end{document}